\documentclass[%
 reprint,
superscriptaddress,
 amsmath,amssymb,
 aps,
]{revtex4-1}

\usepackage{braket}
\usepackage{graphicx}
\usepackage{dcolumn}
\usepackage{bm}
\usepackage{comment}
\usepackage{amsthm}
\theoremstyle{definition}
\newtheorem{theorem}{Theorem}[section]

\newtheorem*{theorem*}{Definition}

\begin{document}

\preprint{APS/123-QED}

\title{Spatial-Translation-Induced Discrete Time Crystals}

\author{Kaoru Mizuta}
 \email{mizuta.kaoru.65u@st.kyoto-u.ac.jp}
\affiliation{%
 Department of Physics, Kyoto University, Kyoto 606-8502, Japan
}%
\author{Kazuaki Takasan}%
\affiliation{%
 Department of Physics, Kyoto University, Kyoto 606-8502, Japan
}%
\author{Masaya Nakagawa}
 \affiliation{
  RIKEN Center for Emergent Matter Science (CEMS), Wako, Saitama 351-0198, Japan
 }
\author{Norio Kawakami}
\affiliation{%
 Department of Physics, Kyoto University, Kyoto 606-8502, Japan
}%

\date{\today}

\begin{abstract}
A discrete time crystal is a phase unique to nonequilibrium systems, where discrete time translation symmetry is spontaneously broken.  Most of conventional time crystals proposed so far rely on spontaneous breaking of on-site symmetries and their corresponding on-site symmetry operations. In this Letter, we propose a new time crystal dubbed ``spatial-translation-induced discrete time crystal (STI-DTC)'', which is realized by spatial translation and its symmetry breaking. Owing to the properties of spatial translation, in this new time crystal, various time crystal orders can emerge only by changing the filling but not changing the driving protocol. We demonstrate that local transport of charges or spins shows a nontrivial oscillation, enabling detection and applications of time crystal orders, and also provide promising platforms including quantum circuits. Our proposal opens up a new avenue of realizing time crystal orders by spatial translation in various quantum simulators.
\end{abstract}

\pacs{Valid PACS appear here}
\maketitle


\textit{Introduction.}---A system whose Hamiltonian is periodic in time is called Floquet system. The past decade has seen a tremendous growth of interest in such Floquet systems, which produce a variety of phases controlled by periodic driving \cite{Oka09,Kitagawa11,Lindner11,Grushin14,Takasan17A,Takasan17B,Wang13,Jotzu14}. It has also turned out that Floquet systems have novel phases that cannot exist in equilibrium, such as  anomalous Floquet topological phases hosting chiral edge states despite the vanishing Chern numbers \cite{Kitagawa10,Rudner13,Titum16,Po16,Mukherjee17}.

One of the most striking phases in Floquet systems is a discrete time crystal phase. It has been indeed  proved that time crystals, where time translation symmetry is spontaneously broken, cannot exist in thermal equilibrium \cite{Watanabe15}. Thus, time crystals are inherent in nonequilibrium systems. In particular, time crystals realized in Floquet systems, called discrete time crystals (DTCs), are phases where discrete time translation symmetry is spontaneously broken and the resulting oscillation frequency of local observables is robust to perturbations. DTCs have attracted much interest because of their theoretical developments \cite{Else16,Yao17} and experimental realizations in various systems \cite{Choi17,Zhang17,Rovny18A,Rovny18B,Pal18}.

In most of conventional DTCs proposed so far, symmetry operation and phases brought by many-body localization (MBL) or spontaneous symmetry breaking (SSB) are utilized to realize DTC orders \cite{Sacha15,Else16,Yao17,Else17,Gong18,Ho17,Russomanno17,Zeng17}. However, among them, only on-site symmetries by finite groups $\mathbb{Z}_n$ have been focused on, thus leading to a restriction that changing the driving protocol is required to realize different types of DTC orders. 

In this Letter, focusing on spatial translation symmetry, which is a non-local but \textit{infinite} group symmetry, we propose new DTCs realized by spatial translation and its symmetry breaking, and we thereby provide a feasible platform to realize various kinds of time crystals. First, note that spatial translation symmetry breaking can induce various orders. For example, a variety of charge density wave (CDW) orders can be realized by changing the filling when discrete spatial translation symmetry is spontaneously broken. Therefore, by utilizing this characteristics, in the new DTC, which is dubbed ``spatial-translation-induced DTC''(STI-DTC), various DTC orders can be realized and controlled without changing the protocol in sharp contrast to the previously proposed DTCs. We further demonstrate that, in STI-DTCs, spatial translation induces local transport which shows nontrivial oscillation due to the time crystal orders. This property is characteristic of STI-DTCs, having merits for detection and application of DTC orders. We also provide a general scheme to implement STI-DTCs with quantum circuits. With the above novel properties, STI-DTCs will open up a new way to realizing time crystals in various quantum simulators and to their application to quantum information processing.

\textit{Definition and example of DTCs.}---First of all, let us clarify  the definition of DTCs. Originally, time crystals are introduced as systems where spontaneous time translation symmetry breaking (TTSB) occurs \cite{Wilczek12}. Since we consider only the cases when the Hamiltonian is periodic in time with period $T$, this means that local observables have a period different from the Hamiltonian's period $T$ in its (quasi)steady states. However, the definition characterized only by TTSB is inadequate because trivial examples such as Rabi oscillation are included.  To preclude such examples and define DTC as a stable phase of matter,  it should be defined as a phase where not only TTSB occurs but also the period of oscillation is robust to perturbations which do not change the driving period \cite{Yao17}. When the period of the local observables is $nT$, the phase is called $nT$-DTC.

DTCs are realized in several ways \cite{Else16,Else17,Iadecola17,Khemani16}. Among them, we  focus on the one which relies on SSB and its corresponding symmetry operation. Assume that Floquet operator $U_f$, which is the time evolution operator of one period under the time-periodic Hamiltonian $H(t)$, is written in the form of
\begin{equation}\label{XiDT}
U_f  \equiv \mathcal{T} \exp \left\{ -i \int_0^T H(t) dt \right\} = X \exp (-iDT),
\end{equation}
where the effective Hamiltonian $D$ induces spontaneous $\mathbb{Z}_n$-symmetry breaking and $X$ is the corresponding symmetry operation. Then TTSB occurs. Intuitively, this is because some Floquet eigenstates become cat states which are superpositions of $n$ symmetry-broken ordered states. Since the quasienergies of such cat states  are equidistantly separated by $2\pi/nT$, the oscillation of the cat states with the period $T$ becomes unstable and instead the superpositions thereof (i.e. macroscopically stable ordered states) show the nontrivial oscillation.
Robustness of TTSB behavior is supported by several ways such as prethermalization and MBL \cite{Else17,Keyserlingk16}.  In both cases, the Floquet operator is unitarily equivalent to the form of Eq. (\ref{XiDT}) even if there is a small perturbation, and hence TTSB behavior is robust.

\begin{figure}[t]
\begin{center}
    \includegraphics[height=2.75cm, width=8.5cm, clip]{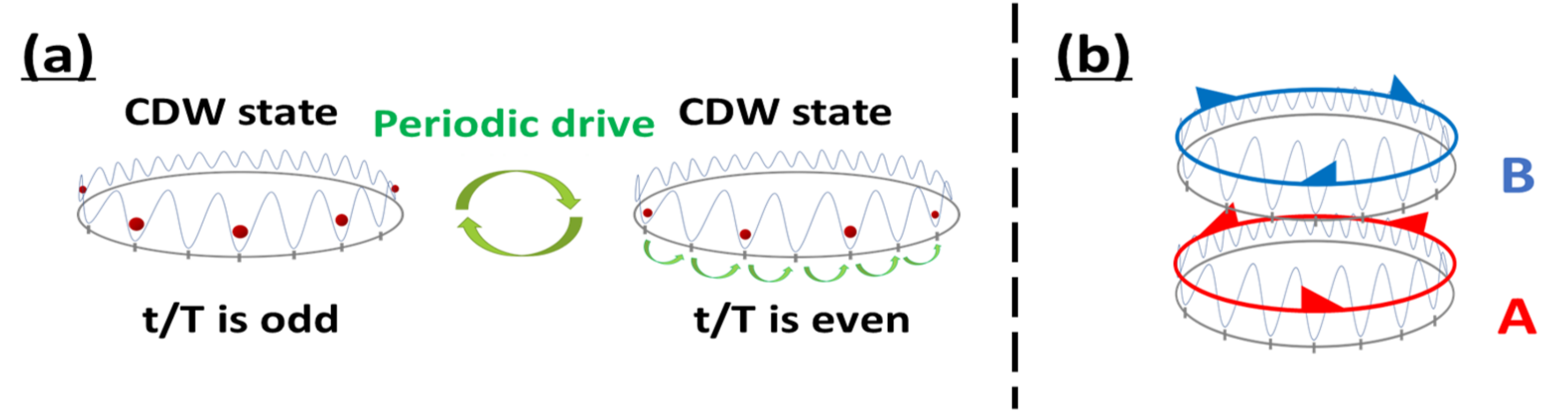}
    \caption{(a) Intuitive picture of the STI-DTCs. If spatial translation is realized, the particle number oscillates with  a period of $nT$. (b) The model for STI-DTCs. During one period, particles on the lower ring are translated anticlockwise and those on the upper one are translated clockwise, thus in total, there is no pumping.}
    \label{ConventionalNew}
  \end{center}
\end{figure}

\textit{Spatial-translation-induced DTC.}---Here, we propose a new type of time crystals : spatial-translation-induced DTCs (STI-DTCs), where the symmetry operation $X$ is spatial translation and $D$ shows spontaneous spatial translation symmetry breaking. 

Let us consider a spinless fermion system in a one-dimensional ring. We consider a lattice system at half-filling. If the spinless fermions have long-range repulsive interactions and the temperature of the initial state is low enough, spatial translation symmetry breaking occurs and one of the two symmetry-broken states realizes (Fig. \ref{ConventionalNew} (a)). In each state, one fermion localizes in every two sites and forms a CDW state. Suppose that spatial translation operation by one site can be realized by a certain periodic driving.  Then, the dynamics of this system can be described as Fig. \ref{ConventionalNew} (a).  Since positions of localized fermions change every period, the particle density at a certain site oscillates with a double period of the driving. The nontrivial $2T$-period oscillation is expected to be stabilized by the CDW order. Therefore, this system would be a $2T$-DTC if the assumption were correct \cite{Equivalence}.

Compared with conventional DTCs, it is notable that STI-DTCs can utilize $\mathbb{Z}_n$ orders brought by CDW for any integer $n(\geq2)$ to realize DTC orders. In other words, $nT$-DTCs are expected to be realized at $1/n$-filling since a particle localizes in every $n$ sites in each of CDW states. Thus, STI-DTCs can realize various DTC orders with the same protocol only by changing the filling.

One important question is how the spatial translation operation can be realized by local Hamiltonians. If any long time can be taken for one period, spatial translation is possible in a one-dimensional ring by Thouless pumping \cite{Thouless83}, which is adiabatically performed. It has already been experimentally realized in cold atoms \cite{Nakajima16,Lohse16,Schweizer16} and the combination with CDW has been theoretically suggested \cite{Nakagawa17,Zeng15,Zeng16,Taddia17}. However, Thouless pumping, which requires infinite time even for one period, is not suitable for realization of DTCs. To overcome this difficulty, we propose below a one-dimensional ladder ring as a candidate of STI-DTCs, which is nonadiabatically realizable (Fig. \ref{ConventionalNew} (b)). 

\begin{figure}[t]
\begin{center}
          \includegraphics[height=5cm, width=8.5cm, clip]{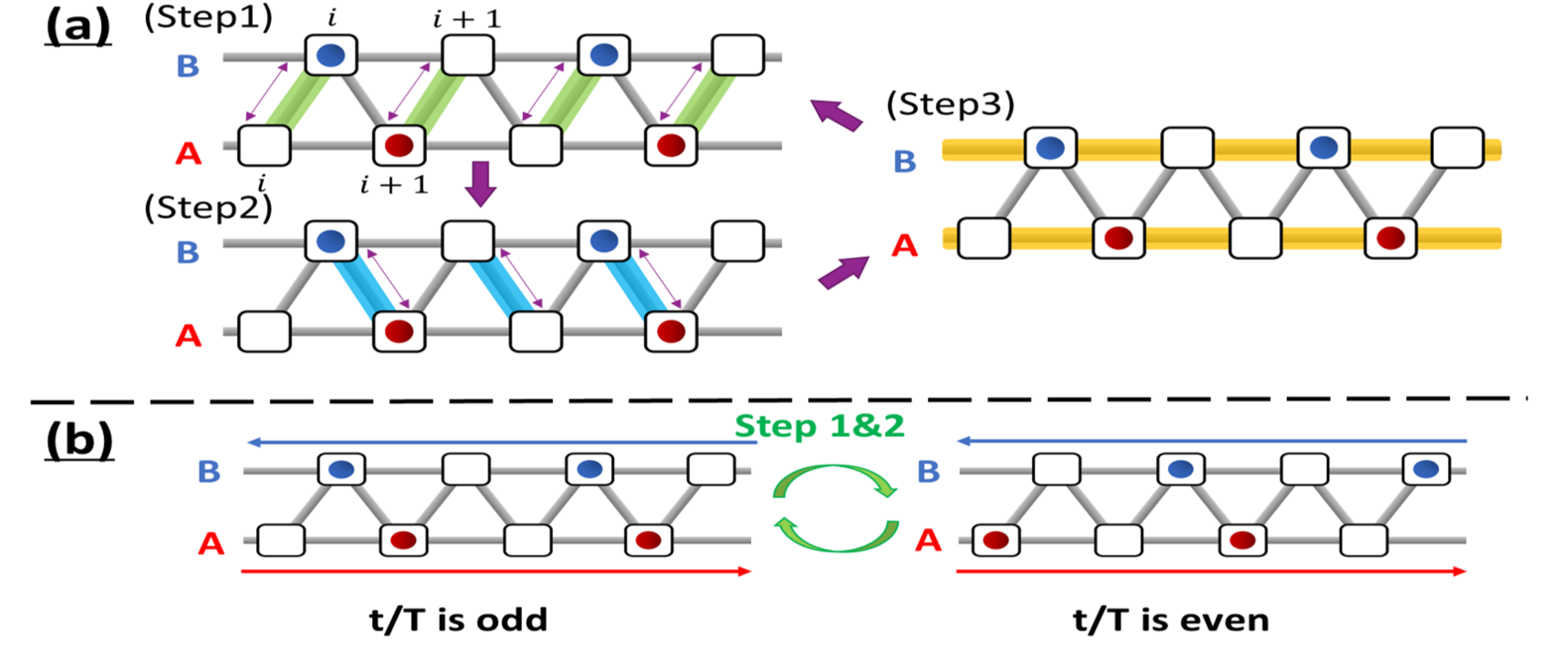}
           \caption{(a) Hamiltonian described by Eq. (\ref{Hamiltonian}) for one period. After Step1 or Step2, the particle numbers are exchanged between sites linked by green or blue lines. Yellow lines in Step3 mean repulsive interaction. (b) Dynamics of the model. The period of particle number becomes $2T$ at half-filling.}
           \label{HamiltonianDynamics}
        \end{center}
\end{figure}

\textit{Model in 1D ladder.}---Here, we describe how to realize a STI-DTC in a one-dimensional ladder ring. In this model, as shown in Fig. \ref{ConventionalNew} (b),  spatial translation $\mathbb{T}_A$ by one site in the sublattice A and the opposite one $\mathbb{T}^{-1}_B$ in the sublattice B are induced every period. Since the total amount of pumping is zero, the time-dependent Hamiltonian is nonadiabatically realizable by switching local Hamiltonians as follows \cite{Po16} (Fig. \ref{HamiltonianDynamics} (a)):

\begin{equation}\label{Hamiltonian}
H(t) = \begin{cases}
H_1 & (0 \leq t \leq \tau/2)  \\
H_2 & (\tau/2 < t \leq \tau) \\
H_{\mathrm{SSB}} & (\tau < t \leq T),
\end{cases}
\end{equation}
where each Hamiltonian is defined as 
\begin{eqnarray}
H_1 &=& - \frac{\pi}{\tau}\sum_i (c_{i,A}^\dagger c_{i,B} + h.c.), \label{Hamiltonian1} \\
H_2 &=& - \frac{\pi}{\tau}\sum_i (c_{i+1,A}^\dagger c_{i,B} + h.c.), \\
H_{\mathrm{SSB}} &=& \sum_{\alpha=A,B} \sum_{i,j} \frac{U_{ij}}{2} n_{i,\alpha} n_{j,\alpha}.
\end{eqnarray}
Here, $c_{i,\alpha}$ and $n_{i,\alpha}$ respectively represent the annihilation and the number operator of spinless fermions at site $i$ in a sublattice $\alpha=A,B$. $U_{ij}$ represents strength of long-range repulsive interaction and then $H_{\mathrm{SSB}}$ is a Hamiltonian that induces spontaneous spatial translational symmetry breaking at low temperature.
As seen below, the time evolution under the Hamiltonians $H_1$ and $H_2$ generates the spatial translation $\mathbb{T}_A \otimes \mathbb{T}_B^{-1}$ and the time evolution under  $H_{\mathrm{SSB}}$ stabilizes CDW states.

\textit{Time translation symmetry breaking.}---Let us confirm that TTSB occurs in the system. How the spatial translation $\mathbb{T}_A \otimes \mathbb{T}_B^{-1}$ is generated by the Hamiltonian is explained intuitively here. The detailed calculation is provided in Supplemental Materials \cite{Supplemental}.

First, let us consider the time evolution under the Hamiltonian $H_1$ given in Eq. (\ref{Hamiltonian1}), which induces hopping between sites $(i,A)$ and $(i,B)$ for every $i$. Since the duration of imposing $H_1$ is fine-tuned, a particle at $(i,A)$ is completely transferred to $(i,B)$ and vice versa by  the time evolution under $H_1$. Similarly, by  the time evolution under $H_2$, exchanges of particles occur between sites $(i+1,A)$ and $(i,B)$ (Fig. \ref{HamiltonianDynamics} (a)). When we consider the dynamics under $H_1$ and $H_2$, a particle at $(i,A)$ moves to $(i,B)$, and after that, it reaches $(i+1,A)$. On the other hand, a particle at $(i,B)$ moves to $(i-1,B)$. Therefore, the spatial translation by one site in A and the opposite translation in B are realized by $H_1$ and $H_2$, as described by
\begin{equation}\label{U2U1}
  e^{-iH_2 \tau/2} e^{-iH_1 \tau/2}= (\mathbb{T}_A \otimes \mathbb{T}_B^{-1}) \, \, U_p,
\end{equation}
where $U_p \equiv \exp \left\{ -i\pi \sum_i n_{i,A} (n_{i,B}+n_{i+1,B}) \right\}$ is the phase which stems from the commutation relation of fermion operators \cite{Supplemental}. Here, a global phase other than $U_p$ is removed by a proper gauge transformation.

Then, from Eq. (\ref{U2U1}), the Floquet operator $U_f$ is described as follows:
\begin{equation}\label{FloquetOp}
U_f = (\mathbb{T}_A \otimes \mathbb{T}_B^{-1}) \, \, \exp (-iH_{\mathrm{SSB}}(T-\tau)) \, \, U_p.
\end{equation}
Though its form is slightly different from Eq. (\ref{XiDT}) by the existence of $U_p$, TTSB can be induced since the spatial translation  $\mathbb{T}_A \otimes \mathbb{T}_B^{-1}$ moves particles regardless of $U_p$. In fact, in the Heisenberg picture,
\begin{equation}
n_{i,A}(mT) = n_{i-m,A}, \quad n_{i,B}(mT) = n_{i+m,B}
\end{equation}
is satisfied for $m\in \mathbb{N}$ \cite{Supplemental}. Thus, when a CDW state, where spatial translation symmetry is broken, is prepared as the initial state, then  the particle density or the current at each site oscillates with a period different from the Hamiltonian (Fig. \ref{HamiltonianDynamics} (b)). At $1/n$-filling, since particles localize at every $n$ sites, TTSB occurs and $nT$-oscillation is observed.

\begin{figure*}
\hspace{-1cm}
\begin{center}
    \includegraphics[height=6.5cm, width=18cm]{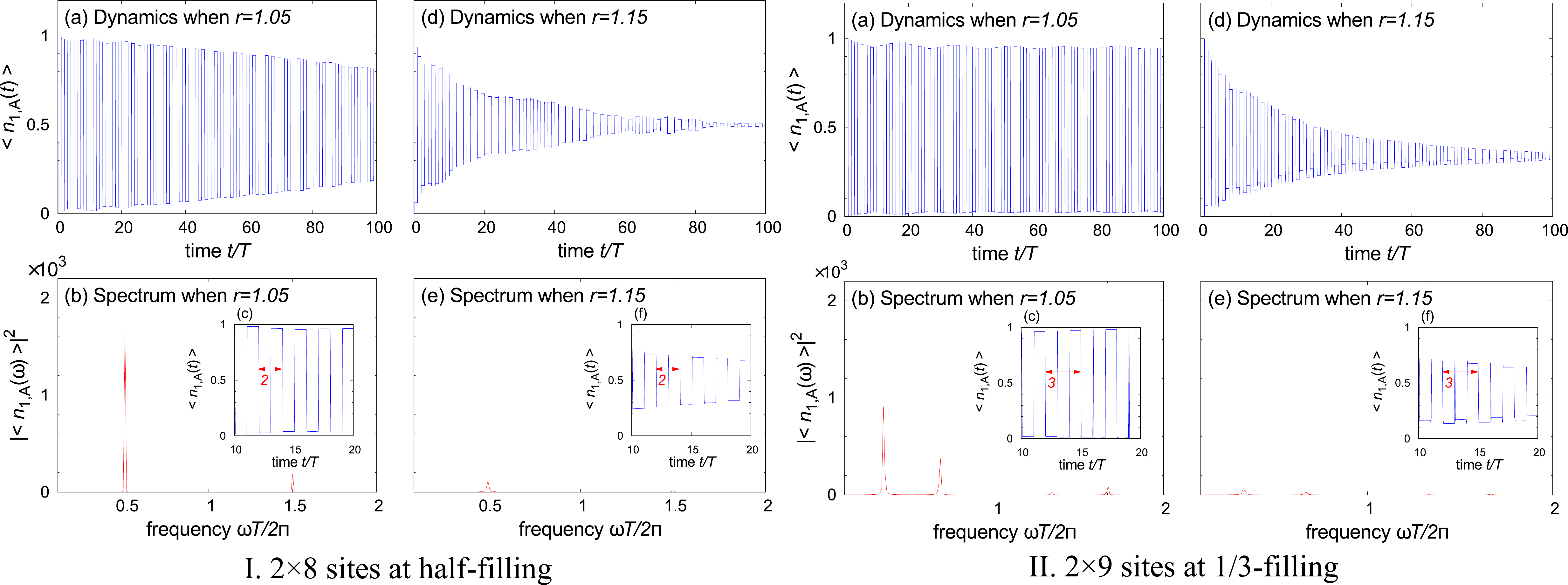}
    \caption{Dynamics of the particle number and its Fourier spectrum in the case of I. $2\times8$ sites at half-filling and in the case of II. $2\times9$ sites at 1/3-filling, (a,b) when $r=1.05$ and (d,e) when $r=1.15$, respectively. (c) and (f) are enlarged figures of (a) and (d), respectively. In each calculation, the interaction $U_{\mathrm{int}}T_{\mathrm{SSB}}$ is 1.0 and the duration $\tau/T$ is 0.1. The peaks at $\omega T/2\pi =0$ in the spectrum are neglected since they are not related to TTSB behavior. In each situation, the dynamics shown in the figures is the full time evolutions but not the stroboscopic ones.}
    \label{RobustCDW}
  \end{center}
\end{figure*}

\textit{Robustness of TTSB.}---One of the nontrivial properties of DTCs is robustness, that is, the oscillation frequency of the observables is hardly influenced by some small perturbations. In the case of $nT$-DTCs, the peak structure of the Fourier component of the oscillation at $\omega T = 2\pi/n$ does not move nor split even if there is a small perturbation \cite{Yao17}. 

To confirm the robustness of DTCs, let us consider perturbations on symmetry operations \cite{Zhang17,Gong18,Huang17}. In the case of STI-DTCs, the perturbation is assumed to be on the spatial translation operation, that is, on the duration of $H_2$. Then, the perturbed Hamiltonian $H(t)$ is described as follows:
\begin{equation}
H(t) = \begin{cases}
H_1 & (0 \leq t \leq \tau/2) \\
H_2 & (\tau/2 < t \leq (1+r)\tau/2) \\
H_{\mathrm{SSB}} & ((1+r)\tau/2 < t \leq T).
\end{cases}
\end{equation}
When $r = 1$, Eq. (9) is reduced to the unperturbed case described by Eq. (2). Thus, the value $|r-1|$ represents the strength of the perturbation. The independent parameters of the system are $r$ and $U_{\mathrm{int}}T_{\mathrm{SSB}}$ where $T_{\mathrm{SSB}}\equiv T-(1+r)\tau/2$ represents the duration of $H_{\mathrm{SSB}}$. Since the theorem about prethermalization in \cite{Else17} is not necessarily applicable to the current system \cite{Prethermalization}, robustness to the perturbation is examined by the exact diagonalization for finite systems. We assume a long-range repulsive interaction 
$U_{ij}=U_{\mathrm{int}}/r_{ij}^3$, where $r_{ij}$ and $U_{\mathrm{int}}$ represents the distance between sites $i$ and $j$, and the strength of the interaction respectively. Such a long-range interaction is realized in trapped ions and diamond NV centers, which are platforms of DTCs \cite{Zhang17,Choi17}.

Figure \ref{RobustCDW} represents the results when $U_{\mathrm{int}}T_{\mathrm{SSB}}=1.0$ in the case of I. $2\times8$ sites at half-filling and II. $2\times9$ sites at 1/3-filling. Here the initial state is assumed to be an equilibrium state under $H_{SSB}$ at low temperature, which spontaneously breaks the spatial translation symmetry. In this calculation, the initial temperature is zero, and the odd sites in both sublattices are occupied in the initial state \cite{Initial}. Supplemental Materials provide the case with finite initial temperature, which shows a similar result \cite{Supplemental}. In both cases, when $r=1.05$,  the oscillation of the particle number hardly decays (See (a)), thus TTSB behavior is robust. On the other hand, when $r=1.15$, the oscillation rapidly decays and then DTC order is lost (See (d)). Robustness can be examined also from their Fourier spectrum described by (b) and (e). Each of the peaks at $\omega T/2\pi=1/2$ and $\omega T/2\pi=1/3$ corresponds to each of $2T$-oscillation and $3T$-oscillation. It is notable that these peaks do not move from their original positions when $r=1.05$. This property is unique to DTCs. 

\begin{figure}[t]
\begin{center}
\includegraphics[height=3.5cm, width=8.5cm]{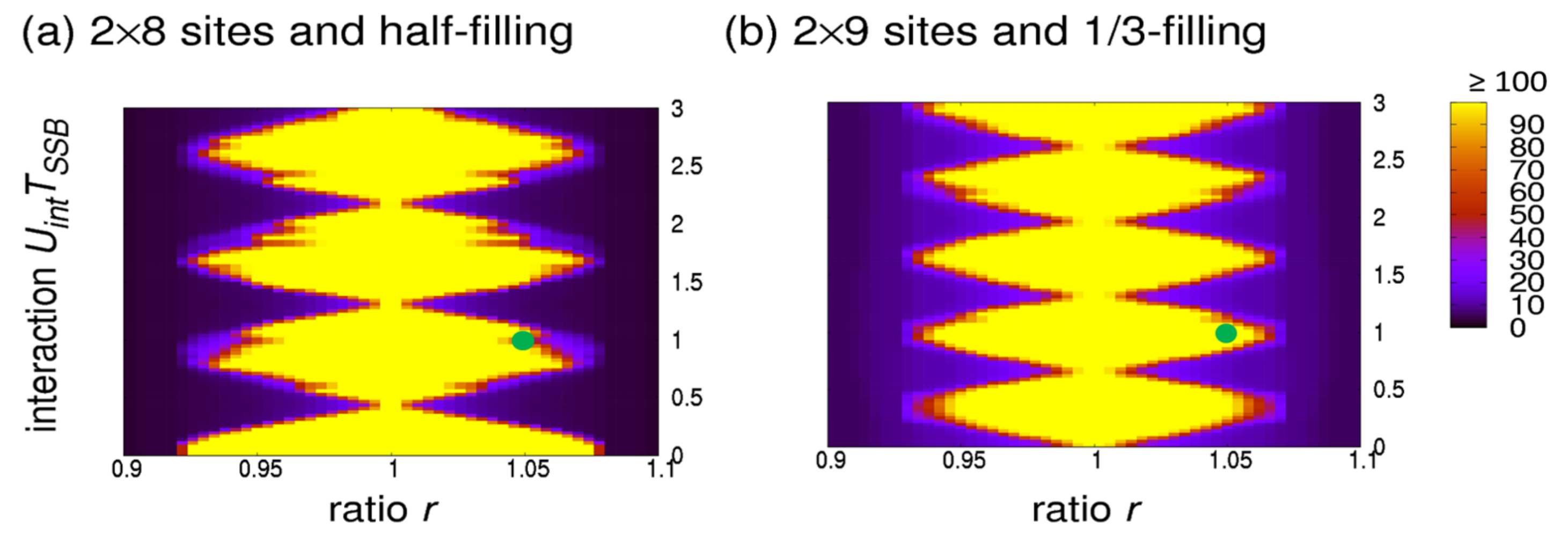}
\caption{Lifetime for each $r$ and $U_{\mathrm{int}}T_{\mathrm{SSB}}$ is described by the colors. In the yellow regions, lifetime is more than $100T$, where $T$ represents the driving period. The green points in (a) and (b) correspond to the cases described in Fig. \ref{RobustCDW}. I. (a) and  Fig. \ref{RobustCDW}. II. (a), respectively.}
\label{CDWPD}
\end{center}
\end{figure}

From these results, in both cases I and II of Fig. 3, there seem to be thresholds, where the DTC orders are lost, between $r=1.05$ and $r=1.15$. To examine the existence of the thresholds, lifetime of STI-DTCs is also calculated for each $r$ and $U_{\mathrm{int}}T_{\mathrm{SSB}}$ (Fig. \ref{CDWPD}). Here, the lifetime is defined as the time when the amplitude of the oscillation becomes $90\%$ of the initial value. Yellow regions indicate that the lifetime is more than $100T$, thus DTC order is stable there. On the other hand, the black regions represent the cases where no robust oscillation is observed, and the boundary represents the thresholds where DTC orders are lost. Thus, Fig.4 (a) and (b) can be regarded as phase diagrams of $2T$-DTCs and $3T$-DTCs, respectively.

Note that robustness is observed at half-filling even without the interaction (Fig. \ref{CDWPD} (a)). This phenomenon originates from the statistics of fermions. In the unperturbed case, where the Floquet operator is described by Eq. (\ref{FloquetOp}), the effective density-density interaction appears in $U_p$ because of the commutation relation of fermion operators. Even in perturbed cases, though the form is different from $U_p$, the effective density-density interaction appears and could stabilize CDW states at half-filling.

\textit{Experimental setup.}---STI-DTCs are expected to be realized in various systems since the essential ingredients for them are spatial translation and its symmetry breaking. For example, spin systems can also realize STI-DTCs by utilizing SWAP gates and antiferromagnetic order \cite{Supplemental}. Thus it is expected that there are various platforms for realizing STI-DTCs.

For example, in trapped ions and Rydberg atoms, long-range interactions which can realize CDW order have been experimentally observed \cite{Zhang17,Bernien17}. On the other hand, in cold atoms, selective hopping by $H_1$ and $H_2$ is theoretically proposed by moving optical lattices \cite{Po16} and long-range interactions can be realized by dipole-dipole interactions and electric fields \cite{Huang17}. Thus, these are candidates for STI-DTCs.

One of the most promising platforms is a quantum circuit as a quantum simulator \cite{Nielsen,IBM}. Since dynamics of Floquet systems is described by $U_f$, STI-DTCs can be realized once the unitary gate of $U_f$ is prepared.  In fact, $U_f$ described by Eq. (\ref{FloquetOp}) is realizable since the time evolution operators under $H_1$,$H_2$ and $H_\mathrm{SSB}$ can be composed of NOT gates, SWAP gates and CPHASE gates \cite{Supplemental}. Quantum circuits are or will be realized by various platforms such as superconducting qubits and quantum dots \cite{IBM,Georgescu14,Wendin17,Hensgens17}. Thus STI-DTCs on quantum circuits will be realized by various platforms.

\textit{Conclusions.}---In this paper, a new type of DTCs: spatial-translation-induced DTCs have been proposed. In this time crystal, spatial translation operation changes the system from a CDW state to another CDW state, thus the particle number  oscillates with a period different from its Hamiltonian. Unlike almost all the conventional DTCs, DTC orders in this system can be controlled by changing its filling. One-dimensional ladder ring under periodic drive has been proposed as a model of STI-DTCs which is realizable without adiabaticity.

One question left open is to clarify the origin of robustness. In this paper, robustness of the STI-DTCs has been confirmed by numerical calculations. Despite inapplicability of the theorem about prethermalization in \cite{Else17}, robustness of TTSB behavior has been observed, as seen in other examples \cite{Choi17,Ho17,Machado17,Huang17}. Though it has been qualitatively demonstrated in this paper that the CDW order supports the robustness, its quantitative evaluation by analytical calculation is desired.

\begin{acknowledgments}
We would like to thank M. Ueda, Z. Gong, R. Hamazaki, and H. Katsura for fruitful discussions. This work is supported by a Grant-in-Aid for Scientific
Research on Innovative Areas ``Topological Materials Science''
(KAKENHI Grant No. JP15H05855) and also JSPS KAKENHI
(Grants No. JP16J05078, No. JP16K05501, and No. JP18H01140). K.T. thanks JSPS for support from a Research Fellowship for Young Scientists. M.N. is supported by RIKEN Special Postdoctoral Reseacher Program.
\end{acknowledgments}

\providecommand{\noopsort}[1]{}\providecommand{\singleletter}[1]{#1}%

\clearpage

\renewcommand{\thesection}{S\arabic{section}}
\setcounter{section}{0}
\renewcommand{\theequation}{S\arabic{equation}}
\setcounter{equation}{0}
\renewcommand{\thefigure}{S\arabic{figure}}
\setcounter{figure}{0}

\onecolumngrid
\begin{center}
{\large {\bfseries Supplemental Materials for \\ "Spatial-Translation-Induced Discrete Time Crystals"}}
\end{center}
\vspace{10pt}
\twocolumngrid

\section{Detailed Calculation for TTSB}
\subsection{Floquet operator}
Assume that the Hamiltonian is given by
\begin{equation}\label{Hamiltonian1SM}
H(t) = \begin{cases}
H_1 & (0 \leq t \leq \tau/2) \\
H_2 & (\tau/2 < t \leq \tau) \\
H_{\mathrm{SSB}} & (\tau < t \leq T)
\end{cases}
\end{equation}
and
\begin{eqnarray}
H_1 &=& - \frac{\pi}{\tau}\sum_i (c_{i,A}^\dagger c_{i,B} + h.c.), \\
H_2 &=& - \frac{\pi}{\tau}\sum_i (c_{i+1,A}^\dagger c_{i,B} + h.c.), \\
H_{\mathrm{SSB}} &=& \sum_{\alpha=A,B} \sum_{i,j} \frac{U_{ij}}{2} n_{i,\alpha} n_{j,\alpha}. \label{CDWHamiltonian}
\end{eqnarray}
Then, we show that the Floquet operator $U_f$ is written as follows,
\begin{eqnarray}\label{FloquetOpSM}
U_f &=& (\mathbb{T}_A \otimes \mathbb{T}_B^{-1}) \, \, \exp (-iH_{\mathrm{SSB}}(T-\tau)) \, \, U_p, \\
U_p &=& \exp \left\{ -i\pi \sum_i n_{i,A} (n_{i,B}+n_{i+1,B}) \right\}.
\end{eqnarray}

First, let us consider $U_1 \equiv \exp (-iH_1\tau/2)$,
\begin{eqnarray*}
U_1  &=& \exp \left( i\frac{\pi}{2}\sum_i (c_{i,A}^\dagger c_{i,B} + h.c.) \right) \\
&=& \prod_i \left\{ \sum_{n=0}^\infty \frac{1}{n!} \left( i\frac{\pi}{2} \right)^n (c_{i,A}^\dagger c_{i,B} + h.c.)^n \right\}.
\end{eqnarray*}
When the integer $n$ is larger than 1 and even, 
\begin{eqnarray*}
(c_{i,A}^\dagger c_{i,B} &+& h.c.)^n \\ 
&=& (c_{i,A}^\dagger c_{i,B} c_{i,B}^\dagger c_{i,A})^{n/2} + (c_{i,B}^\dagger c_{i,A} c_{i,A}^\dagger c_{i,B})^{n/2} \\
&=& n_{i,A}^{n/2} (1-n_{i,B})^{n/2} + n_{i,B}^{n/2} (1-n_{i,A})^{n/2} \\
&=& n_{i,A} (1-n_{i,B}) + n_{i,B} (1-n_{i,A}) \\
&\equiv& P_{i,A}^{i,B}.
\end{eqnarray*}
The operator $P_{i,A}^{i,B}$ is the projection onto the subspace where $n_{i,A}+n_{i,B}=1$. On the other hand, when $n$ is odd, 
\begin{eqnarray*}
(c_{i,A}^\dagger c_{i,B} + h.c.)^n &=& (c_{i,A}^\dagger c_{i,B} + h.c.) (c_{i,A}^\dagger c_{i,B} + h.c.)^{n-1} \\
&=& (c_{i,A}^\dagger c_{i,B} + h.c.) P_{i,A}^{i,B}.
\end{eqnarray*}
Therefore, we obtain $U_1$ as follows,
\begin{equation}
U_1 =  \prod_i \left\{ Q_{i,A}^{i,B}+iP_{i,A}^{i,B} (c_{i,A}^\dagger c_{i,B} + h.c.) \right\},
\label{U1effect}
\end{equation}
where $Q_{i,A}^{i,B} \equiv 1-P_{i,A}^{i,B}$ is the projection onto the subspace where  $n_{i,A}+n_{i,B}$ is 0 or 2. If the Fock states $\{ \ket{\{n\}} \equiv \ket{n_{1,A}...n_{N,A}n_{1,B}...n_{N,B}} \}$ are chosen as the basis, 
\begin{equation}
U_1 \ket{\{n\}} = i^{S(\{n\})} \ket{n_{1,B}...n_{N,B}n_{1,A}...n_{N,A}},
\end{equation}
\begin{equation}
S(\{n\}) = \sum_i \left\{  n_{i,A} (1-n_{i,B}) + n_{i,B} (1-n_{i,A}) \right\},
\end{equation}
where $S(\{n\})$ represents how many times particles are transferred between sites $(i,A)$ and $(i,B)$ whose particle numbers are different. From these equations, $U_1$ gives not only an exchange of particles between $(i,A)$ and $(i,B)$ but also the corresponding phase $ i^{S(\{n\})}$. Similarly, $U_2 \equiv \exp (-iH_2\tau/2)$ exchanges the particle numbers between sites $(i+1,A)$ and $(i,B)$ for every $i$ and gives the phase $i^{T(\{n\})}$, where
\begin{equation}\label{Tn}
T(\{n\}) = \sum_i \left\{  n_{i+1,A} (1-n_{i,B}) + n_{i,B} (1-n_{i+1,A}) \right\}.
\end{equation}
Therefore, $U_2 U_1$ is calculated as follows (Note that $U_2$ is performed after $U_1$, thus $T(\{n\})$ appears in the form where indices $(i,A)$ and $(i,B)$ are exchanged for every $i$ ) :
\begin{equation}
U_2 U_1 =(\mathbb{T}_A \otimes \mathbb{T}_B^{-1}) i^{U},
\end{equation}
where $U$ is given as 
\begin{eqnarray*} 
U &=& \sum_i \{ n_{i,A} (1-n_{i,B}) + n_{i,B} (1-n_{i,A}) \\ 
&\qquad& + n_{i+1,B} (1-n_{i,A}) + n_{i,A} (1-n_{i+1,B}) \} \\
&=& 2 \sum_i ( n_{i,A}+n_{i,B} ) \\
&\qquad& -2\sum_i n_{i,A}(n_{i,B}+n_{i+1,B}).
\end{eqnarray*}
Since $\sum_i ( n_{i,A}+n_{i,B} )$ is conserved, this term merely gives a global phase to the state. By removing this term by a proper gauge transformation, the phase term $i^U$ is derived as follows:
\begin{equation}
i^U \simeq \exp \left\{ -i\pi \sum_i n_{i,A}(n_{i,B}+n_{i+1,B}) \right\} = U_p.
\end{equation}
Since the Hamiltonian $H_{\mathrm{SSB}}$ commutes with $\mathbb{T}_A \otimes \mathbb{T}_B^{-1}$, we obtain the Floquet operator $U_f$ as Eq. (\ref{FloquetOpSM}).

\subsection{TTSB}
When STI-DTCs are composed of spinless fermions, the additional phase term $U_p$ is included in the Floquet operator, which is different from the previous studies. Here, we would like to confirm that the Floquet operator $U_f$ given by Eq. (\ref{FloquetOpSM}) induces TTSB, even if it includes $U_p$.

For any integer $n (\geq1)$, the particle number on $(i,A)$ at $t=nT$, $n_{i,A}(nT)$ is given as follows in the Heisenberg picture,
\begin{equation}
n_{i,A}(nT) = U_f^{-n} n_{i,A} U_f^n.
\end{equation}
In the case of $n=1$, since $U_{\mathrm{SSB}} \equiv \exp (-iH_{\mathrm{SSB}}(T-\tau))$ and $U_p$ commute with $n_{j,\alpha}$ for any $j$ and $\alpha=A,B$,
\begin{eqnarray}
n_{i,A}(T) &=& U_p^{-1} U_{\mathrm{SSB}}^{-1} n_{i-1,A} U_{\mathrm{SSB}} U_p  \notag \\
&=& n_{i-1,A}
\end{eqnarray}
is satisfied. By repeating this calculation, we have
\begin{equation} \label{TransA}
n_{i,A}(nT) = n_{i-n,A} 
\end{equation}
for any integer $n\geq1$. Similarly, the equation for the sublattice B is given by 
\begin{equation} \label{TransB}
n_{i,B}(nT) = n_{i+n,B}.
\end{equation}
Therefore, if the initial state is prepared as a CDW state, the particle number at each site oscillates with the corresponding period, and then TTSB occurs.

\section{Realization in spin systems}
\subsection{Model and TTSB}\label{SpinModel}
In the main text, it has been proved that spontaneous TTSB can occur in spinless fermion systems. In order to explore experimental setups for STI-DTCs, it is important to clarify whether or not they are realizable in other systems. We would like to suggest that STI-DTCs are also realizable in spin systems. Assume that the system is a one-dimensional ladder ring and each site has a spin half. The essential point is to prepare spatial translation $\mathbb{T}_A \otimes \mathbb{T}_B^{-1}$ and spatial translation symmetry breaking in spin systems. Therefore, the time-dependent Hamiltonian $H(t)$ is assumed to be the form of Eq. (\ref{Hamiltonian1SM}), where each Hamiltonian is defined as
\begin{eqnarray}
H_1 &=& \frac{\pi}{2\tau} \sum_i (1+\vec{\sigma}_{i,A} \cdot \vec{\sigma}_{i,B}) \label{SpinH1}, \\
H_2 &=& \frac{\pi}{2\tau} \sum_i (1+\vec{\sigma}_{i+1,A} \cdot \vec{\sigma}_{i,B}) \label{SpinH2}, \\
H_{\mathrm{SSB}} &=& J_{\mathrm{int}}\sum_{\alpha=A,B}\sum_i \sigma_{i,\alpha}^z \sigma_{i+1,\alpha}^z. \label{SpinHssb}
\end{eqnarray}
The time evolution operators under the Hamiltonians $H_1$ and $H_2$ are calculated as
\begin{equation}
 e^{-iH_1\tau/2} = \prod_i \chi_{(i,A)}^{(i,B)}, \quad e^{-iH_2\tau/2} = \prod_i \chi_{(i+1,A)}^{(i,B)}.
\end{equation}
Here, a global phase is removed by a gauge transformation. The operator $\chi_\gamma^\delta$ is defined by
\begin{equation}
\chi_\gamma^\delta \equiv \frac{1}{2}(1+ \vec{\sigma}_\gamma \cdot \vec{\sigma}_\delta),
\end{equation}
and exchanges the states between the site $\gamma$ and the site $\delta$, that is, 
\begin{equation}
\chi_\gamma^\delta \ket{\psi_1}_\gamma \ket{\psi_2}_\delta = \ket{\psi_2}_\gamma \ket{\psi_1}_\delta.
\end{equation}
Thus, $\chi_\gamma^\delta$ is called a SWAP gate. From the same reason as the spinless fermion systems, the time evolution under $H_1$ and $H_2$ generates spatial translation in the sublattice A and oppsite translation in the sublattice B, thus the Floquet operator $U_f$ is written as
\begin{equation}
U_f \equiv \mathcal{T} \exp \left\{ -i \int_0^T H(t) dt \right\} = (\mathbb{T}_A \otimes \mathbb{T}_B^{-1}) \, \, e^{-iH_{\mathrm{SSB}}(T-\tau)}.
\end{equation}
Therefore, if we prepare an antiferromagnetic state as the initial state, then spontaneous TTSB is induced by this Floquet operator.

\subsection{Robustness}\label{SpinRobust}

\begin{figure}[t]
\begin{center}
\includegraphics[height=6cm, width=8.5cm]{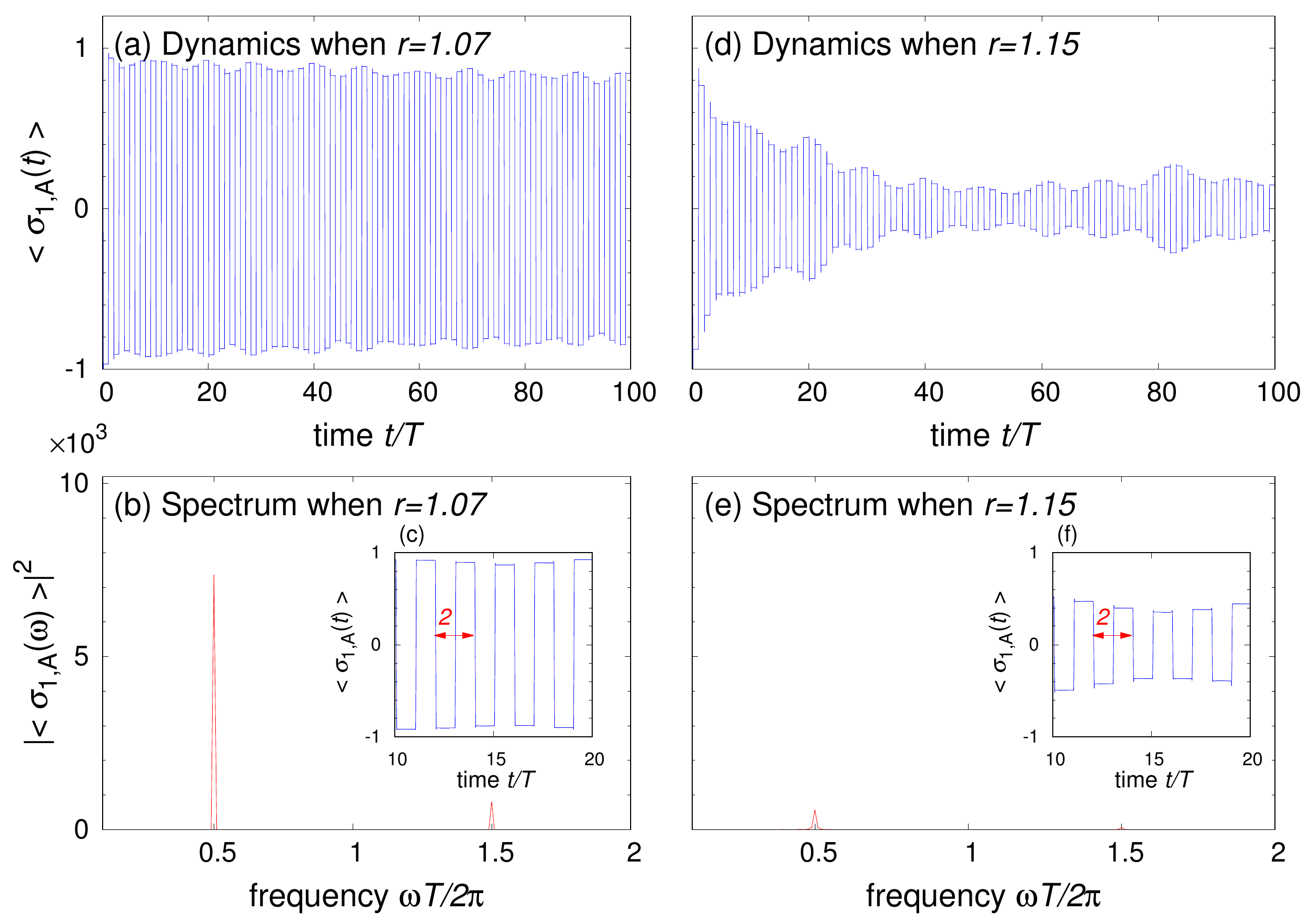}
\caption{Dynamics of the $z$ component of spin at a certain site and its Fourier spectrum (a,b) when $r=1.07$ and (d,e) when $r=1.15$. (c) and (f) are enlarged figures of (a) and (d), respectively. In each calculation, the interaction $J_{\mathrm{int}}T_{\mathrm{SSB}}$ is 0.1. }
\label{SpinDTC}
\end{center}
\end{figure}

\begin{figure*}
\begin{center}
\begin{tabular}{c}
\begin{minipage}{0.5\hsize}
        \begin{center}
          (i) Spin systems
          \includegraphics[height=5cm, width=8.5cm]{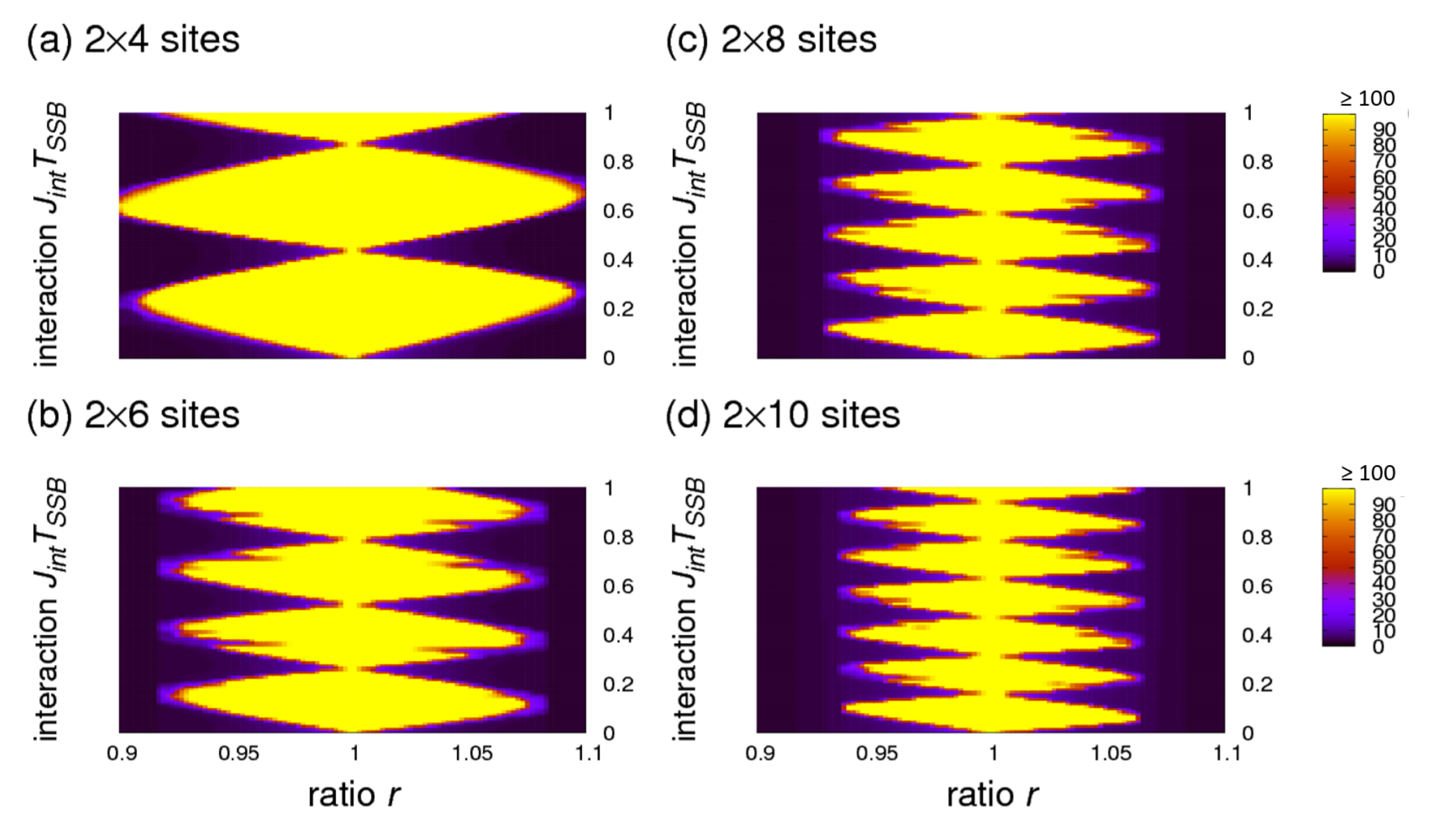}
        \end{center}
      \end{minipage}

      \begin{minipage}{0.5\hsize}
        \begin{center}
        (ii) Spinless fermion systems
          \includegraphics[height=5cm, width=8.5cm]{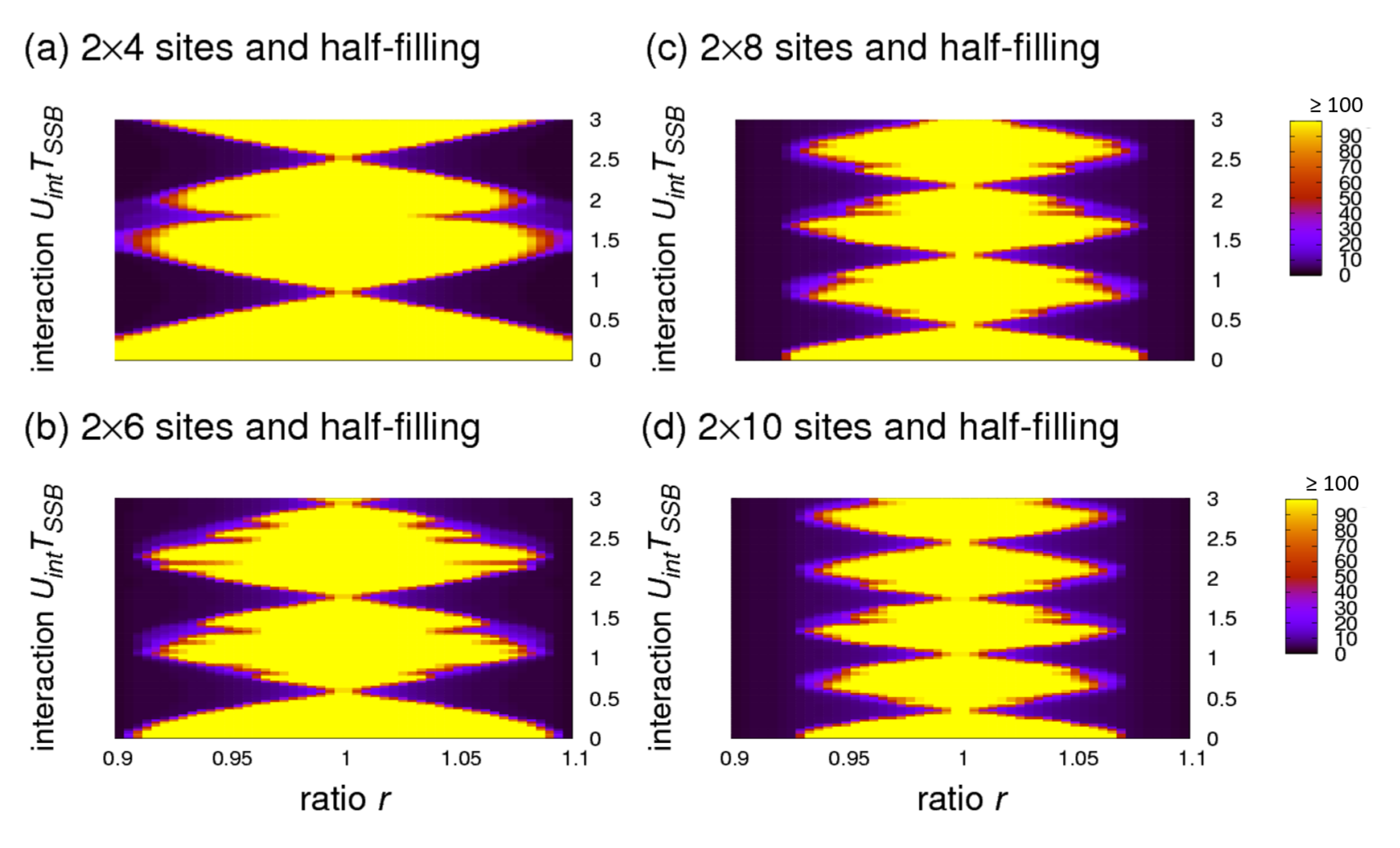}
        \end{center}
      \end{minipage}

      \end{tabular}
    \caption{Size-dependence of the lifetimes (i) in spin systems and (ii) in spinless fermion systems at half-filling. In both cases, the numbers of sites are (a) $2\times4$, (b) $2\times6$, (c) $2\times8$, and (d) $2\times10$, respectively. The lifetime for each $r$ and each parameter of the interaction is described by the
colors. In the yellow regions, lifetime is more than $100T$, where $T$ represents the driving period.}
    \label{SizeDependence}
  \end{center}
\end{figure*}

In this section, we show that the STI-DTC in spin systems is also robust to small perturbations. Assume that the perturbed Hamiltonian is described by
\begin{equation}
H(t) = \begin{cases}
H_1 & (0 \leq t \leq \tau/2) \\
H_2 & (\tau/2 < t \leq (1+r)\tau/2) \\
H_{\mathrm{SSB}} & ((1+r)\tau/2 < t \leq T),
\end{cases}
\end{equation}
where $H_1,H_2$ and $H_{\mathrm{SSB}}$ are given by Eqs. (\ref{SpinH1}),(\ref{SpinH2}), and (\ref{SpinHssb}). Since SSB under the Hamiltonian $H_{\mathrm{SSB}}$ occurs only at zero temperature, the initial state is assumed to be the ground state of $H_{\mathrm{SSB}}$, that is, the antiferromagnetic state. Since independent parameters of the system are $r$ and $J_{\mathrm{int}}T_{\mathrm{SSB}}$, the expectation value of $\sigma_{1,A}^z$ is calculated for each $r$ and  $J_{\mathrm{int}}T_{\mathrm{SSB}}$ by exact diagonalization.

Figure \ref{SpinDTC} shows the time evolution of the expectation value of $\sigma_{i,A}^z$ (See (a),(c),(d), and (f)) and its Fourier spectrum (See (b) and (d)). All of them are calculated under $J_{\mathrm{int}}T_{\mathrm{SSB}}=0.1$ and the number of sites $N=2\times8$. When $r=1.07$, as shown in Fig. \ref{SpinDTC} (a),(b), and (c), the oscillation hardly decays and the highest peak of the Fourier spectrum is pronounced and does not move from $\omega T = \pi$.  Thus, the time crystal order is maintained in that case. Figure \ref{SpinDTC} (d),(e), and (f) shows the strongly perturbed case $r=1.15$. Then, the oscillation rapidly decays and the highest peak at $\omega T=\pi$ becomes much lower. The strong perturbation $r=1.15$ breaks time crystal order.

The lifetime for each $r$ and $J_{\mathrm{int}}T_{\mathrm{SSB}}$ in $2\times8$ sites is shown in Fig. \ref{SizeDependence} (i) (c). The definition of lifetime is the same as $2T$-DTCs composed of spinless fermions, which has been described in the main text. Lifetime is identified by colors in the graph, and the yellow regions represent the cases when the lifetime is more than 100 periods. Thus, it is concluded that time crystal order is stable within these yellow regions.

\section{Size-dependence of TTSB behavior}

\begin{figure}
\begin{center}

          \includegraphics[height=6cm, width=8.5cm]{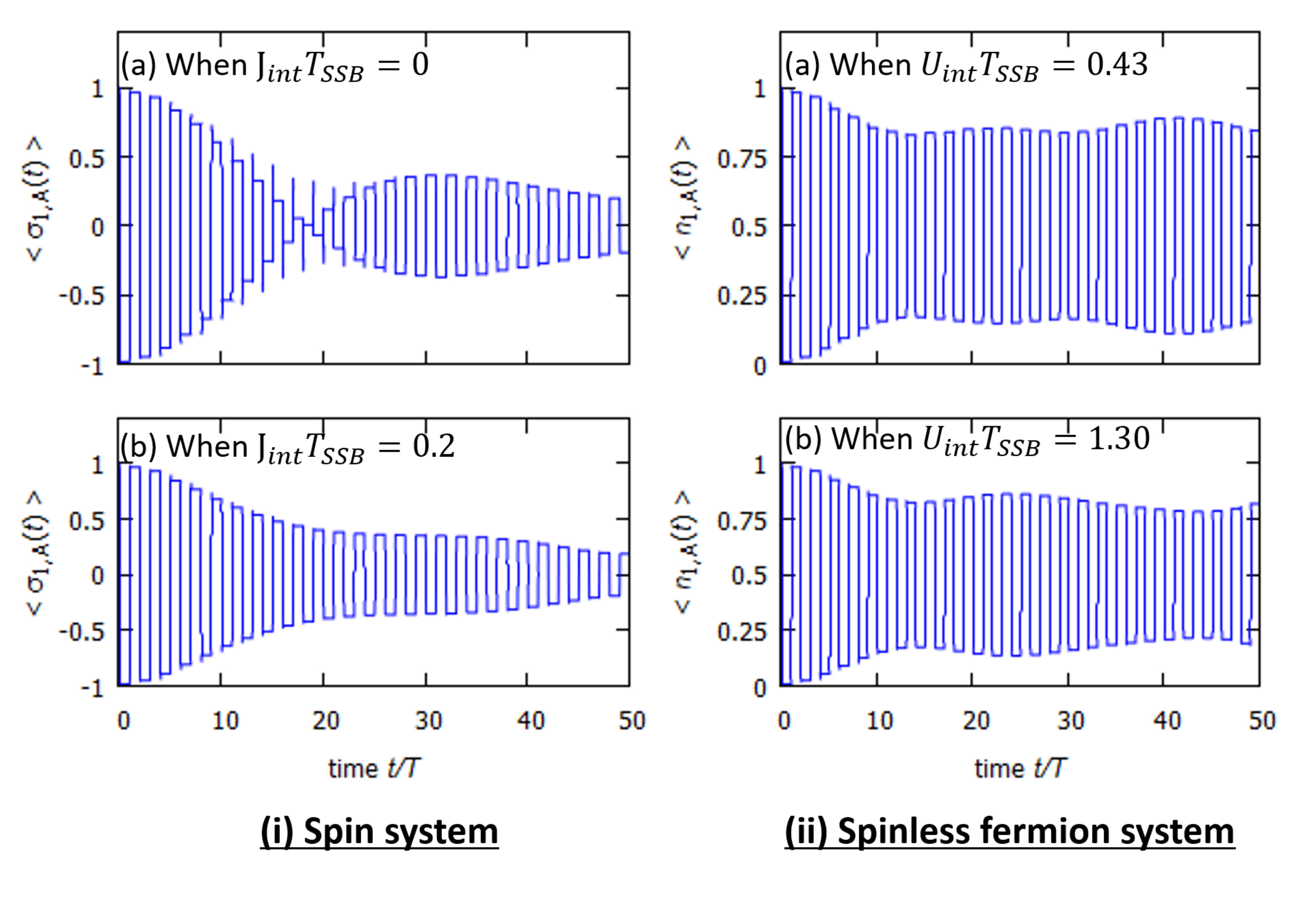}
           \caption{Full dynamics at different nodes in the case of (i) \\ spin systems and (ii) half-filled spinless fermion systems \\ under $L=8, r=1.05$. In both cases, the behaviors in the \\ early time regime $t\leq10T$ are similar to each other.}
           \label{ResonantDynamics}
   
  \end{center}
\end{figure}

In the main text, the robustness of TTSB behavior has been confirmed for a certain finite system size. It is important to examine the size-dependence of this robustness. The lifetimes for different size systems are also calculated by exact diagonalization, as shown in Fig. \ref{SizeDependence}. Figure \ref{SizeDependence} (i) represents the case of spin systems, and Fig. \ref{SizeDependence} (ii) represents the case of half-filled spinless fermion systems. In both  cases, the lifetimes look periodic in the strength of the interaction and there are ``nodes'', which are points at $r=1$ where the lifetime dramatically decreases even when $r$ is slightly moved from 1. Moreover, the number of nodes is proportional to the number of sites in each sublattice $L$. This comes from the resonance between the energy from $H_{\mathrm{SSB}}$ and the driving frequency $2\pi/T_{\mathrm{SSB}}$. Let us consider the case of spin systems. From Fig. \ref{SizeDependence} (i), the nodes appear at 
\begin{equation}\label{resonant}
J_{\mathrm{int}}T_{\mathrm{SSB}} = \frac{n\pi}{2L}, \quad n \in \mathbb{Z}.
\end{equation}
The period of the lifetime in $J_{\mathrm{int}}T_{\mathrm{SSB}}$ is equal to $\pi/2L$. The ground state of $H_{\mathrm{SSB}}$ is prepared at $t=0$, thus the effect of $\exp(-iH_{\mathrm{SSB}}T_{\mathrm{SSB}})$ is approximated by $\exp(-iE_GT_{\mathrm{SSB}})$ in an early time regime, where  the ground state energy $E_G$ of $H_{\mathrm{SSB}}$ is equal to $-2LJ_{\mathrm{int}}$. When $J_{\mathrm{int}}T_{\mathrm{SSB}}$ differs by $n\pi/2L$, the change brought by this difference in the Floquet operator $U_f$ is equal to $\exp(in\pi)=\pm1$, which is expected to be trivial. Therefore, the lifetime is periodic in $J_{\mathrm{int}}T_{\mathrm{SSB}}$, and since there is no robustness in the case of no interaction, the nodes appear when Eq. (\ref{resonant}) is satisfied. Resonant behavior of STI-DTCs can be also seen from the dynamics at different nodes (Fig. \ref{ResonantDynamics} (a) and (b)).  In an early time regime, in which the effect of perturbations is small yet, the dynamics when the interaction is fine-tuned by Eq. (\ref{resonant}) are similar to one another.

In the case of spinless fermion systems, since the nontrivial phase term $U_p$ appears and the interaction is long-ranged, conditions for appearance of nodes cannot be simply described as Eq.  (\ref{resonant}). It is also notable that there is robustness even when there is no interaction because of the effective density-density interaction brought by $U_p$. However, similar resonance takes place because similarity of the dynamics at different nodes is observed, as shown in Fig. \ref{ResonantDynamics} (c) and (d). Such resonance that spoils robustness when there is a long-range interaction is also observed in \cite{Ho17SM}. 

In any size examined here, robustness of TTSB behavior exists. Since the promising experimental resources such as cold atoms, trapped ions, and quantum circuits, have at most $O(10^1)$ sites, we can conclude from the results that robustness of STI-DTCs will be observed in these setups. As a matter of theoretical interest, the behavior in the thermodynamic limit is greatly important, but this problem is left for future work.

\section{TTSB for finite-temperature initial states}
In the main text, TTSB and its robustness are confirmed in the case of the zero-temperature initial state defined by $H_{\mathrm{SSB}}$.  In this appendix, we would like to show that robustness also exists in the case of finite but low temperatures and that their is no necessity of fine-tuning of the initial state to the zero-temperature ground state.

\subsection{Mean-field analysis of the initial state}

\begin{figure*}
\begin{center}
\begin{tabular}{c}

\hspace{-0.5cm}
\begin{minipage}{0.5\hsize}
        \begin{center}
          \includegraphics[height=6cm, width=9cm]{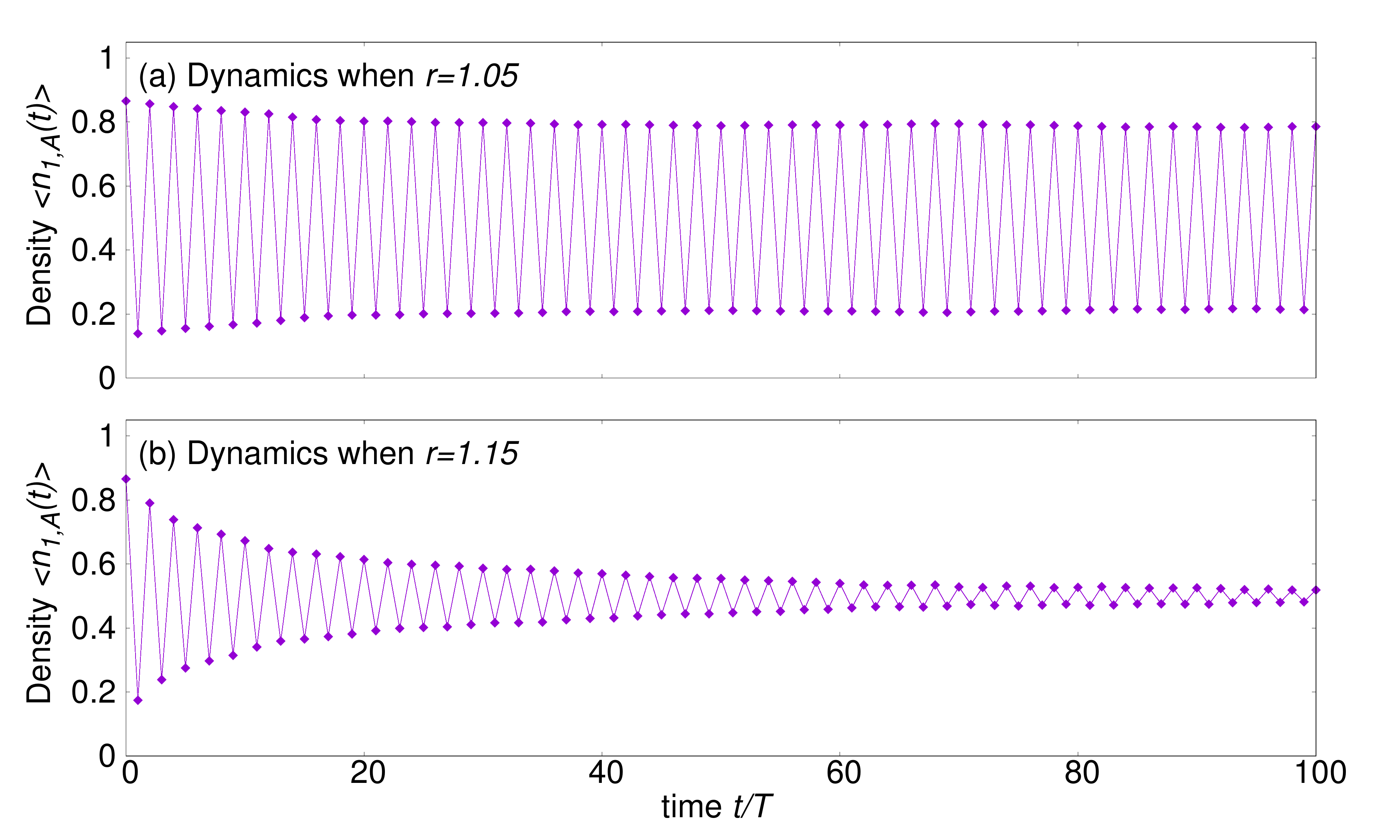}
          \caption{The stroboscopic dynamics at $t/T \in \mathbb{Z}$ of the particle density at the site (1,A) when the initial state is a symmetry-breaking ordered state at finite temperature. (a) When $r=1.05$, the oscillation survives and the DTC order is robust. (b) When $r=1.15$, the perturbation is strong and destroys the DTC order.}
          \label{FiniteTemp}
        \end{center} 
      \end{minipage}      
\hspace{0.5cm}
\begin{minipage}{0.5\hsize}
        \begin{center}
          \includegraphics[height=6.0cm, width=9cm]{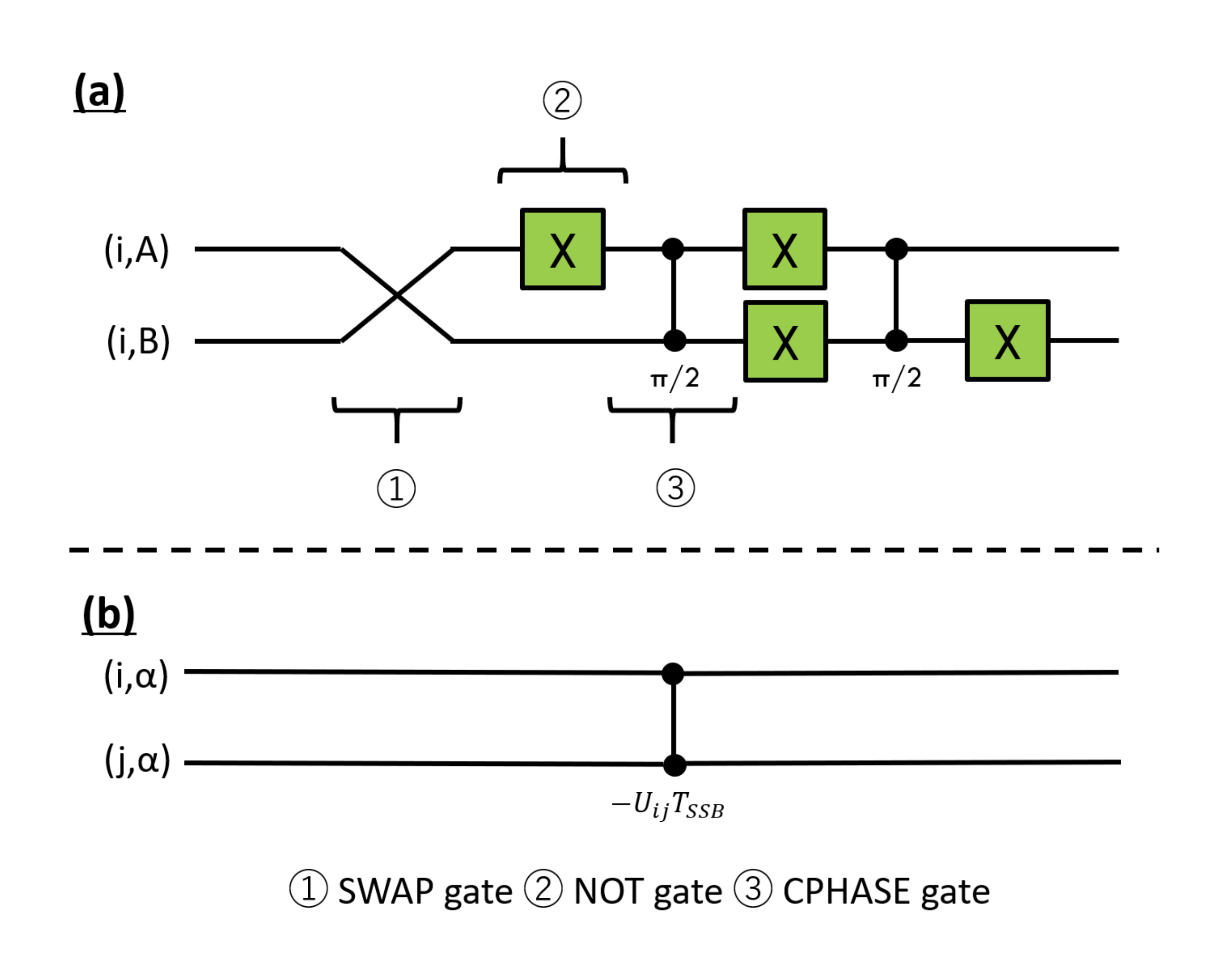}
          \caption{(a) The gate which realizes $U_1$. The gates \textcircled{\footnotesize1}, \textcircled{\footnotesize2}, and \textcircled{\footnotesize3} are a SWAP gate, a NOT gate, and a Controlled-PHASE (CPHASE) gate, respectively. A CPHASE gate gives a certain phase only to the state $\ket{11}$. (b) The gate which realizes $U_3$. It is just a CPHASE gate which gives the phase of $-U_{ij}T_{\mathrm{SSB}}$ to $\ket{11}$. }
          \label{Gate}
        \end{center}
      \end{minipage}

      \end{tabular}
  \end{center}
\end{figure*}

First, the initial state is specified by a mean-field analysis when the inverse temperature $\beta$ under $H_{\mathrm{SSB}}$ is finite. Here the discussion is limited to the case of half-filling, but generalization to $1/n$-filling with generic $n$ is straightforward. Let us assume that SSB under the equilibration by $H_{\mathrm{SSB}}$ takes place and the initial state has a $\mathbb{Z}_2$ spatial order. Then the mean field is described by
\begin{equation}\label{meanfield}
\left< n_{i,\alpha} \right> = \frac{1+(-1)^{i-1}m}{2},
\end{equation}
where $m\in[-1,1]$ represents the mean field. Since the sublattices A and B are independent with each other, only one of the sublattices $\alpha=$ A or B is focused on. By approximating the interaction in $H_{\mathrm{SSB}}$ by the mean field, $H_{\mathrm{SSB}}$ is replaced by the symmetry-breaking Hamiltonian $H_{\mathrm{b}}$, where

\begin{equation}
H_{\mathrm{b}} = \sum_{i,j\neq i} \frac{1}{2} \frac{U_{\mathrm{int}}}{|i-j|^3} \frac{1+(-1)^{j-1}m}{2} n_{i,\alpha}.
\end{equation}
When defining $\tilde{\sigma}_{i,\alpha}$ by $\tilde{\sigma}_{i,\alpha}\equiv (-1)^{i-1}(2n_{i\alpha}-1)$ and using the conservation of particle number at half-filling described by $\sum_i (-1)^i \tilde{\sigma}_{i,\alpha}=0$, we obtain

\begin{eqnarray}
H_{\mathrm{b}} &=& -U_{\mathrm{int}}m\gamma \sum_{i} \tilde{\sigma}_{i,\alpha} + \mathrm{Const.}, \\
\gamma &\equiv& - \frac{1}{4} \sum_{k=1}^{k=N/2} \frac{(-1)^k}{k^3} + O(1/N^3).
\end{eqnarray}

The self-consistent equation for the mean field $m$ at the inverse temperature $\beta$ is given as follows:
\begin{equation} \label{selfconsistent1}
m = \left< \tilde{\sigma}_{i,\alpha} \right> = \mathrm{Tr} \left[ \tilde{\sigma}_{i,\alpha} \mathrm{e}^{-\beta H_\mathrm{b}} \right]/ \mathrm{Tr} \left[ \mathrm{e}^{-\beta H_\mathrm{b}}\right].
\end{equation}
Here, the trace $\mathrm{Tr}$ means taking a sum over states in which half of the sites are occupied. By calculating Eq. (\ref{selfconsistent1}), we obtain the self-consistent equation as follows:
\begin{equation} \label{selfconsistent2}
m=\frac{1}{N} \frac{\sum_{n=0}^{N/2} (4n-N)( _{N/2}C_n)^2 \mathrm{e}^{4\beta\gamma m U_{\mathrm{int}}n}}{\sum_{n=0}^{N/2} ( _{N/2}C_n)^2 \mathrm{e}^{4\beta\gamma m U_{\mathrm{int}}n}},
\end{equation}
where $ _nC_r \equiv n!/\{r!(n-r)!\}$ is the binomial coefficient. In the zero-temperature limit ($\beta \to \infty$), the solutions of Eq. (\ref{selfconsistent2}) are $m= \pm 1$, which reproduce complete CDW states. If the inequality
\begin{equation} \label{CondSSB}
\beta U_\mathrm{int} > \frac{1}{\gamma} \frac{\sum_{n=0}^{N/2}( _{N/2}C_n)^2}{\sum_{n=0}^{N/2}(4n-N)^2( _{N/2}C_n)^2}
\end{equation}
is satisfied, the self-consistent equation (\ref{selfconsistent2}) has nonzero solutions of $m$, that is, SSB occurs. Thus, if the inverse temperature $\beta$ which satisfies Eq. (\ref{CondSSB}) is chosen, the equilibrium state $\rho_\alpha$ in each sublattice $\alpha$ is described by
\begin{equation} \label{stateSSB}
\rho_\alpha = \frac{\sum_{\{n_{i,\alpha}\}} \mathrm{e}^{2\beta U_\mathrm{int} m \gamma \sum_i (-1)^{i-1}n_{i,\alpha}}\ket{\{n_{i,\alpha}\}}\bra{\{n_{i,\alpha}\}}}{\mathrm{Tr} \left[\mathrm{e}^{2\beta U_\mathrm{int} m \gamma \sum_i (-1)^{i-1}n_{i,\alpha}}\right]},
\end{equation}
where the value $m$ is determined by Eq. (\ref{selfconsistent2}). Here, the sum $\sum_{\{n_{i,\alpha}\}}$ and the trace $\mathrm{Tr}$ are taken over states which satisfy the half-filling condition. 

\subsection{Robustness for finite-temperature initial states}

Let us calculate the behavior when the initial state is at finite temperature under $H_{\mathrm{SSB}}$. The inverse temperature $\beta$ is chosen so that Eq. (\ref{CondSSB}) is satisfied and the states in the sublattices A and B are prepared independently. Then, the initial density operator $\rho(0)$ of the system breaks the spatial-translation symmetry, and here it is assumed to be $\rho_A \otimes \rho_B$, where $\rho_\alpha$ $(\alpha=A,B)$ is given by Eq. (\ref{stateSSB}). In the state $\rho_A \otimes \rho_B$, particles localize at odd sites in both sublattices A and B, and this state approaches the initial state in the main text in the limit of $\beta \to \infty$.
 
The dynamics of the system driven by the Hamiltonian of Eq. (9) in the main text with the initial state $\rho(0)=\rho_A \otimes \rho_B$ is calculated by exact diagonalization. The parameters are chosen as $U_{\mathrm{int}}T_{\mathrm{SSB}}=1.0$, which is the same as those of Fig. 3 I in the main text, and $\beta U_{\mathrm{int}}=4.5$, at which Eq. (\ref{CondSSB}) is satisfied and the solution of Eq. (\ref{selfconsistent2}) is $m \simeq 0.87$. The result of the stroboscopic dynamics is shown in Fig. \ref{FiniteTemp} in the case of $r=1.05$ and $r=1.15$, which are the same as the values in Fig. 3. In the case of $r=1.05$ shown in Fig. \ref{FiniteTemp} (a), the oscillation of the particle density at the site (1,A) hardly decays despite the perturbation. Therefore, the TTSB behavior is robust to such a weak perturbation when the initial temperature is finite but low, as well as when the initial temperature is zero. On the other hand, a strong perturbation such as $r=1.15$ makes the oscillation decay rapidly from Fig. \ref{FiniteTemp} (b) , and as a result DTC orders are lost.

From these discussions, STI-DTCs have robustness to  the weak perturbation even when the initial state is at finite temperatures under the Hamiltonian $H_{\mathrm{SSB}}$. Thus, there is no necessity of fine-tuning of the strength of the interaction, the durations of the operations by $H_1$ and $H_2$, and even the initial state to realize DTC orders because of robustness of DTCs.
\section{Realization by Quantum Circuits}

In this section, we describe how to realize STI-DTCs by quantum circuits. In order to realize STI-DTCs, it is enough to prepare the unitary gate equivalent to the Floquet operator $U_f$ since Floquet systems can be simulated by the repetition of the unitary gate $U_f$. In principle, any unitary gate is constructible by unitary operations on 1 qubit and CNOT gates, which is well known as universal quantum computation \cite{NielsenSM}. Here, we describe $U_1 \equiv \exp(-iH_1\tau/2),\, U_2 \equiv \exp(-iH_2\tau/2)$, and $U_3 \equiv \exp (-iH_{\mathrm{SSB}}T_{\mathrm{SSB}})$ concretely in the case of spinless fermion systems.

\subsection*{Unitary gates $U_1$ and $U_2$}
The action of $U_1$, which is given by Eq. (\ref{U1effect}), is to swap particles between sites $(i,A)$ and $(i,B)$ for every $i$ and to give the corresponding phase to the state. The former part is easily realizable by a SWAP gate. The latter part can be completed by combining NOT gates and CPHASE gates since giving the phase $i=e^{i\pi/2}$ only when the total particle number in $(i,A)$ and $(i,B)$ is equal to 1 is required. Therefore, the unitary gate $U_1$ can be constructed by imposing the gates shown in Fig. \ref{Gate} (a)  on sites $(i,A)$ and $(i,B)$ for each $i$. Since the unitary operator $U_2$ acts on sites $(i+1,A)$ and $(i,B)$ for each $i$ in the same way as $U_1$, the unitary gate $U_2$ is also realizable by the gates shown in Fig. \ref{Gate} (a).
\subsection*{Unitary gate $U_3$}
The unitary operator $U_3$, which is equal to $\exp\left( -i\sum_{i,j,\alpha} U_{ij} n_{i,\alpha} n_{j,\alpha}/2 \right)$, represents the effect of the long-range repulsive interaction. Since all of the terms in the summation commute with one another, $U_3$ is realizable by combining unitary gates $U_{ij}^\alpha$ acting on two sites $(i,\alpha)$ and $(j,\alpha)$, where $U_{ij}^\alpha$ is given by 
\begin{equation}
U_{ij}^\alpha = \mathrm{diag}(1,1,1,e^{-iU_{ij}T_{\mathrm{SSB}}}).
\end{equation}
Here, $\{ \ket{0}_{i\alpha}\otimes\ket{0}_{j\alpha}, \ket{0}_{i\alpha}\otimes\ket{1}_{j\alpha}, \ket{1}_{i\alpha}\otimes\ket{0}_{j\alpha} , \ket{1}_{i\alpha}\otimes\ket{1}_{j\alpha}  \}$ is chosen as the basis. Thus, the role of the unitary gate $U_{ij}^\alpha$ is to give a certain phase dependent on the interaction $U_{ij}$ only to the component of $\ket{1}_{i\alpha}\otimes\ket{1}_{j\alpha}$ and this is realizable by a CPHASE gate (Fig. \ref{Gate} (b)).
\nocite{*}
\providecommand{\noopsort}[1]{}\providecommand{\singleletter}[1]{#1}%


\begin{thebibliography}{51}
\bibitem [{Equ()}]{Equivalence}%
  \BibitemOpen
  \href@noop {} {}\bibinfo {note} {If only the stroboscopic dynamics is focused
  on, 2T-DTCs by spatial translation seem equivalent to the conventional DTC in
  spin systems with spin flip \cite{Else17}. However, there is an essential
  difference in that STI-DTCs induce nontrivial currents originating from
  nonlocality of spatial translation.}\BibitemShut {Stop}%
\bibitem [{Pre()}]{Prethermalization}%
  \BibitemOpen
  \href@noop {} {}\bibinfo {note} {In order to apply the theorem,
  $U_{\mathrm{int}}T \ll 1/L$ ($L$ is the number of the sites in each
  sublattice) is at least required. However, driving with the frequency
  proportional to $L$ is difficult, or in the thermodynamic limit, it is
  impossible. Moreover, even in the finite systems, the case when
  $U_{\mathrm{int}}T \simeq 1$ is considered, thus the theorem is not
  applicable.}\BibitemShut {Stop}%
\end{thebibliography}
\end{document}